# Near-deterministic activation of room temperature quantum emitters in hexagonal boron nitride

Nicholas V. Proscia[1,2,#], Zav Shotan[1,#], Harishankar Jayakumar[1], Prithvi Reddy[3], Michael Dollar[1], Audrius Alkauskas[4], Marcus Doherty[3], Carlos A. Meriles[1,2,*] & Vinod M. Menon[1,2,*]

[1]*Dept. of Physics, City College of New York, New York, United States*
[2]*Dept. of Physics, Graduate Center of the City University of New York (CUNY), United States*
[3]*Laser Physics Centre, Research School of Physics and Engineering, Australian National University, Canberra, Australia*
[4]*Center for Physical Science and Technology, Vilnius, Lithuania*

**Applications of quantum science to computing, cryptography and imaging are on their way to becoming key next generation technologies. Owing to the high-speed transmission and exceptional noise properties of photons, quantum photonic architectures are likely to play a central role. A long-standing hurdle, however, has been the realization of robust, device-compatible single photon sources that can be activated and controlled on demand. Here we use strain engineering to create large arrays of quantum emitters in two-dimensional hexagonal boron nitride (hBN). The large energy gap inherent to this Van der Waals material stabilizes the emitters at room temperature within nanoscale regions defined by substrate-induced deformation of the flake. Combining analytical and numerical modeling we show that emitter activation is likely the result of carrier trapping in deformation potential wells localized near the points where the hBN flake reaches the highest curvature. These findings, therefore, hint at novel opportunities for the manipulation of single photon sources through the combined control of strain and external electrostatic potentials under ambient conditions.**

Emerging quantum technologies for cryptography, computing and metrology exploit quantum mechanical effects for enhanced information processing and nanoscale sensing. Though different platform systems are currently being explored, light-based quantum technologies using single-photon emitters as the basic building block are among the frontrunners[1]. Several strategies have been used to realize deterministic single photon sources in the solid state[2], including quantum dots[3], single molecules[4], and point defects in wide bandgap materials such as diamond and silicon carbide[5-9]. Single photon emitters in novel van der Waals materials have garnered recent attention due to their potential for integration with waveguides, microcavities, and other passive components typical in photonic devices. Example 2D systems hosting quantum emitters include $WSe_2$ and $MoS_2$ as well as other transition metal dichalcogenides (TMDs)[10-14].

Here we focus on hexagonal boron nitride (hBN), a wide bandgap semiconductor where defect emission has been shown to be tunable and robust at room temperature[15-22] and above[23]. We use confocal microscopy to investigate the photoluminescence (PL) of point defects within thin hBN flakes deposited on a lithographically patterned $SiO_2$ substrate. Due to Van der Waals forces the flake conforms to the surface topography thus accumulating significant local strain near protruding features. Using large structured arrays of different sizes and geometries we find nearly perfect correspondence between the strained areas of the flake and defect emission. Our modeling supports the notion of defect activation via charge trapping in deformation potential wells. The physics at play has some similarities with that governing the dynamics of excitons in $WSe_2$ monolayers subjected to comparable geometries, as reported recently[24,25]. Unlike TMDs, however, the wide bandgap of hBN can accommodate large potential modulations, sufficient to stabilize the defect charge at room temperature. In particular, we calculate deformation potential wells as deep as 500 meV confined to regions of tensile and compressive strain in the hBN flake that correlate well with the spatial localization of the emitters.

For the first set of experiments we use an array of 155-nm-high nanopillars with diameters ranging from 200 to 700 nm fabricated via electron beam lithography over a large-area silica substrate (Figure 1a); the sample is a commercial, 20-nm-thick flake of hBN grown via chemical vapor deposition (CVD). We follow a wet transfer protocol[26] to drape the flake on the patterned silica substrate (see Methods). This technique takes advantage of the Van der Waals forces to make the flake conform to the surface topography. As an illustration, Figure 1b shows an atomic force microscopy (AFM) image from an hBN fragment where the 20-nm-thick flake folds on itself: We identify single- and double-layer sections near the left and right areas, respectively; bare pillars — visible on the lower, right corner of the image — provide a direct view of the underlying structure. Rather than adopting the geometry of a pole tent canvas, the flake deforms uniformly to reproduce the substrate features regardless the number of layers; the deformation seems to be elastic though hBN piercing was observed for higher pillars (see Methods).

[*] Corresponding authors: cmeriles@ccny.cuny.edu, vmenon@ccny.cuny.edu.

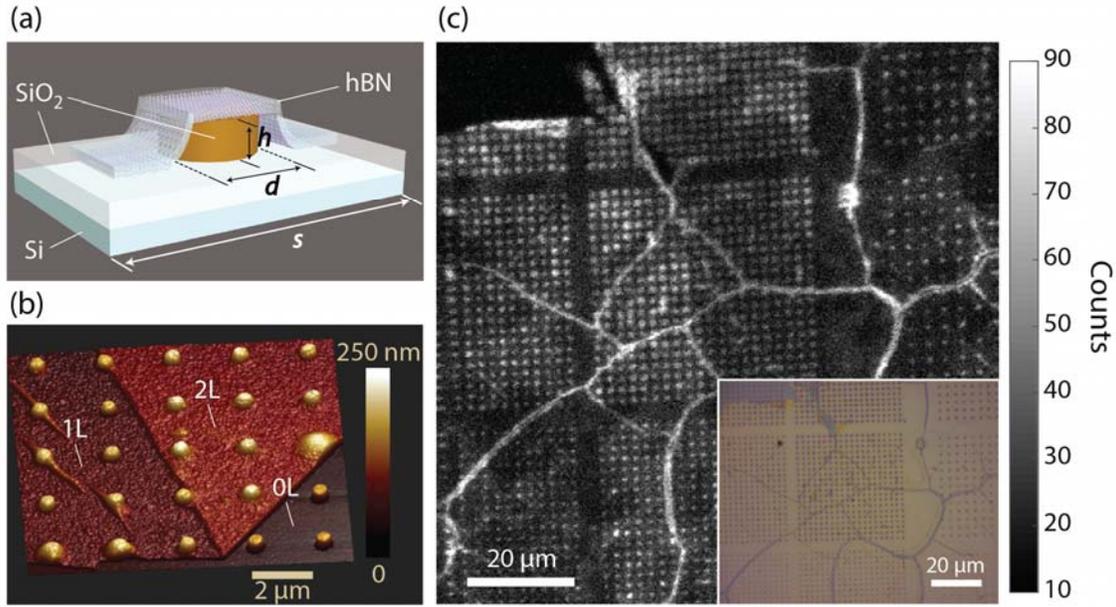

**Figure 1 | Strain-induced activation of single-photon emitters in hBN.** (a) We use a wet transfer protocol to overlay a ~20-nm-thick flake of hBN on a nanostructured silica substrate. For the present experiments, we fabricate an e-beam-defined array of silica nanopillars of variable height $h$, diameter $d$, and spacing $s$. (b) Three-dimensional rendering of an AFM image from folded, ~20-nm-thick hBN. Labels indicate the number of layers, one on the left (1L), and two at the center (2L); bare silica pillars (0L) can be seen on the lower right corner. (c) Room-temperature confocal (main) and optical (inset) images of example nanopillar structures for spacings of 2 μm (left and center arrays) and 3 μm (far right); in all cases the pillar diameter and height are 250 nm and 155 nm, respectively. During confocal scanning, the laser excitation wavelength and intensity are 460 nm and 600 μW/μm², respectively. No fluorescence is detected from areas where the hBN sample is missing (upper left corner in the confocal image).

To investigate the PL of hBN-hosted emitters we use a customized confocal microscope with laser excitation at 460 nm and 488 nm (see Methods). A confocal image of the structure is presented in Figure 1c. Unlike prior work, where visible emitters are created randomly via high-temperature annealing[15], ion irradiation, or surface etching[22], here we find light emission selectively originating from the pillar sites. Comparison with an optical image of the substrate (Figure 1c, inset) shows a seamless correspondence with the underlying structure, independent of the spacing between pillars (compare left and right arrays in the image). Since defects are already present in the hBN flake and can be rendered bright via annealing, (see Supplementary Note 1 and Supplementary Figure 1), our observations suggest a localized defect activation process at the pillar sites making the mechanism nearly deterministic. Fluorescence is also observed along grain boundaries or flake cracks, as reported previously[22]. We collect virtually no emission from sections of the substrate not covered by hBN (upper left corner) and observe a brightness contrast between pillar and inter-pillar sites in excess of 100-fold in most areas. A detailed analysis using literature values for the silicon[27] and hBN[28] indices of refraction shows that the confocal pattern does not originate from pillar-enhanced photon scattering or antenna effects (Supplementary Note 2 and Supplementary Figure 2).

To more thoroughly investigate the PL from the activated sites, we resort to steady-state, time-resolved, and correlation optical spectroscopies. We choose a high-contrast area showing minimal fluorescence in the space between pillars (Figure 2a). Figure 2b shows the spectrum from an example site: Similar to prior reports[15,16,19], we identify well-defined zero-phonon lines (ZPL) accompanied by multiple phonon replicas separated by about 165 meV. The spectrum also displays a broad, near-featureless background with a maximum around 550 nm, possibly the result of spectral diffusion. Lifetime measurement using ultrafast pulses (500 fs, 80 MHz repetition rate) indicate a nearly-exponential fluorescence decay with effective time constant of ~2.5 ns (Figure 2c). Though the number of emitters per site varies, approximately 10% of the pillars show emission from < 3 emitters. As an illustration, Figure 2d presents the results from a pulsed Hanbury-Brown-Twiss (HBT) correlation measurement displaying photon antibunching near zero-delay times. Dividing the area under the central peak by the average area of all other six peaks, we calculate a zero-delay correlation $g^{(2)}(t=0) = 0.50 \pm 0.02$ hinting at the presence of one or two emitters.

Figure 2e presents two complementary examples where we collect the emission spectra from different pillar sites at consecutive times. The set on the left alternates between bright and dark intervals, indicative of a single photon emitter. We assign the zero-phonon line to the peak at 575 nm but the broad background at lower wavelengths suggests strong spectral diffusion on a scale faster than the integration time per trace (1 sec). The system on the right shows a similar behavior though the observation of an intermittent peak at ~625 nm, not correlated with the main peak at 575 nm, hints at the presence of a second photo-active defect. Figure 2f summarizes the statistics from similar measurements at pillar sites featuring a dominant emission peak (approximately 10% of the total): Assigning this peak to the emitter's ZPL, we find a Gaussian-like distribution centered around ~576 nm; the full width at half maximum (FWHM) is ~70 nm, substantially narrower than in prior observations[16,19]. The latter may simply reflect the equivalence between all bright sites, subjected to comparable levels of elastic strain.

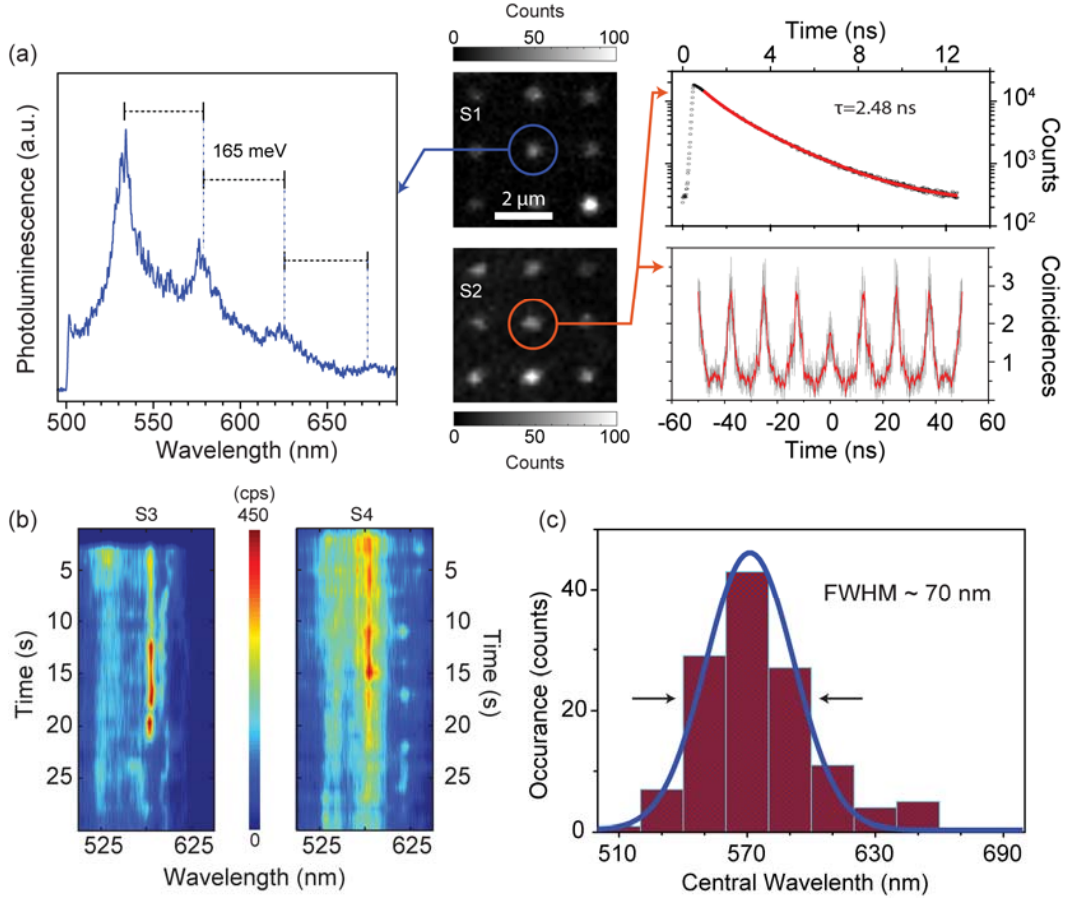

**Figure 2 | Photo-luminescence spectroscopy of strain-activated defects.** (a) Confocal images of the pillar structure along with a representative optical spectrum on the left of a pillar site (blue circle). The relatively sharp zero-phonon line and phonon replicas suggest the emission originates from a single defect. Also shown on the right (top) is the fluorescence lifetime measurement from a neighboring pillar site (red circle). The solid line indicates a fit to a decay of the form $\sim\exp(-(t/\tau)^n)$ with $n=1.2$ and lifetime $\tau = 2.48$ ns. Photon correlation data as determined from pulsed Hanbury-Brown-Twiss measurement of the same pillar site is shown on the right (bottom). By calculating the ratio between the area of the peak at zero time delay and the average area of the other six peaks, we calculate $g^{(2)}(t=0)=0.50\pm0.02$. (b) Photo-luminescence spectra as a function of time for two different pillar sites. The integration time per trace is 1 sec. (c) Probabilistic distribution of the zero-phonon line as determined from multiple pillar sites. In (a) we use pulsed excitation (500 fs, 80 MHz) at 488 nm; in (b), (c) we excite via a cw laser (471 µW) at 460 nm.

To elucidate the interplay between defect activation and local geometry we extend the experiments above to a set of structures of variable shape and size. Two examples are presented in Figure 3, where we study substrates containing large (~2 µm) pillars with circular and triangular cross sections. Similar to the observations in Figure 1, AFM imaging shows that the hBN flake deforms elastically to reproduce the substrate topography. Interestingly, confocal microscopy reveals fairly uniform defect activation along the pillar edges, regardless of the orientation of the hBN lattice relative to the substrate geometry. Further, for the triangular structure we measure light emission of comparable intensity throughout the contour, an intriguing finding because strain at the corners is expected to be substantially larger than at the edges. Given the differing local topologies, this behavior suggests a threshold for the process of defect activation, wherein all defects become bright to optical excitation once a minimum strain threshold is met. As shown in Supplementary Figure 3 for the case of a long, 500-nm-wide ridge, there is virtually no limit in the length of the hBN contour that can be activated.

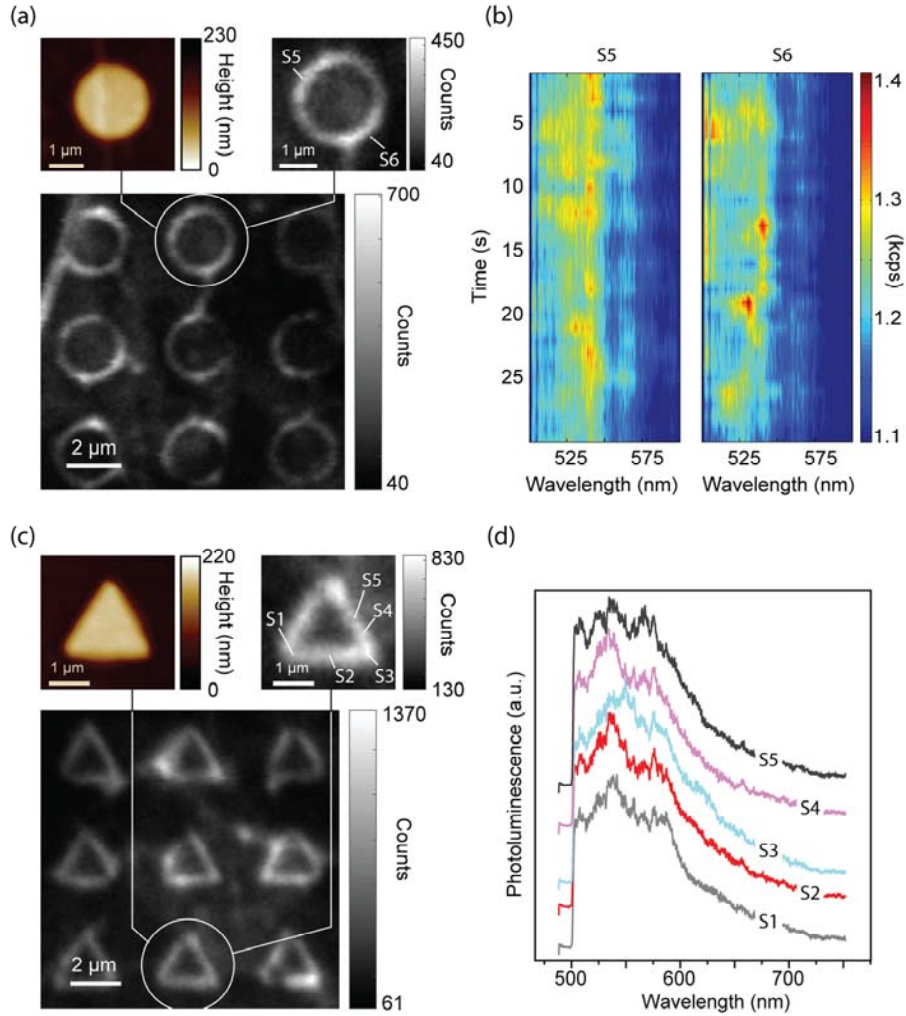

**Figure 3 | Confocal microscopy and micro-spectroscopy of strain-activated emitters in 1D contours.** (a) We transfer a 20-nm-thick hBN flake on a silica substrate featuring 2-μm-diameter pillars; from confocal microscopy (main) we observe preferential emitter activation along the edges of the pillars. The upper inserts show zoomed AFM (left) and confocal (images) of the circled pillar. In the confocal images the integration time per pixel is 2 ms and the laser excitation power and wavelength are 1.7 mW and 460 nm, respectively. (b) Emission spectra as a function of time for sites S5 and S6 along the contour of the circled pillar in (a). (c) Same as in (a) but for pillars with a triangular contour. (d) Integrated emission spectra at sites S7 through S11 along the triangular contour of pillar circled in (a); the integration time is 10 s. Spectra have been displaced vertically for clarity. All silica structures on the substrate in (a) through (d) are 142 nm tall.

The ability to interrogate adjacent but distinguishable (i.e., diffraction-resolved) sites over long perimeters gives us the opportunity to further gauge the role of strain in the observed ZPL dispersion. The multi-peaked structure of most spectra combined with the intermittent blinking of the PL at most locations (Figure 3b) hints at the presence of spectral diffusion between discrete, well-defined configurations of the local charge, consistent with prior observations[19]. Further, integrated spectra from neighboring positions retain common features including the overall structure and center wavelength of the main peaks (spectra in Figure 3d). Rather than transforming abruptly from one location to the next, changes are gradual confirming that strain which varies little over short distances is key to defining the emitters ZPL.

Although the physical nature of the point defects at play in hBN still remains the subject of ongoing research[29], one possible activation route to a bright state entails the capture or loss of one or more charge carriers. Charge localization can take place via various mechanisms. Substrate-induced electrostatic effects is one such mechanism. However we observe similar defect activation on a patterned $Si_3N_4$, which is less electronically active than $SiO_2$, and hence the role of the substrate electrostatics appears to be minimal. Another mechanism of charge confinement arises from a strain-induced potential, which, in turn, can emerge as the combined result of a deformation potential and the piezo-electric effect. Bulk hBN has inversion symmetry and therefore exhibits no piezo-electricity, but weak contributions may still be present in thin flakes if the number of atomic layers is odd. Owing to the trigonal symmetry of the hBN crystal structure, a piezoelectric induced potential in the presence of cylindrical strain must lead to a trigonal distribution of emitters. Since our pillar structures do not display such patterns, we conclude that the piezoelectric induced potential is negligible.

To assess the impact of strain induced deformation potential on charge localization, we first determine from the AFM data the average shape the flake takes in the vicinity of a pillar (Figure 4a). We then make use of the Kirchoff-Love (KL) theory in the limit of thin plates[30-33], and calculate the strain from the second derivative of the deformation relative to the radial distance $r$ to the pillar center; finally, we determine the deformation potential $V_D(r)$ through its proportionality with strain (see Supplementary Note 3 and Supplementary Figure 4). High-resolution atomic force and confocal images are presented in Figures 4a through 4d. Similar to Figure 1, we focus on a small (∅ = 550 nm) cylindrical pillar from the single-layer region of the flake but in this case we consider a site where the number of activated emitters is much larger than one (apparent from the uniform, doughnut-shaped fluorescence image of the pillar, Figure 4d). Using a polynomial fit to reproduce the local topography (Figure 4e and yellow trace in Figure 4f), we determine via the KL theory confining potentials of up to ~500 meV, indicative of robust charge carrier trapping at room temperature. As expected, the radial locations of the extrema are found near the top and base edges of the pillar, where the flake curvature is highest.

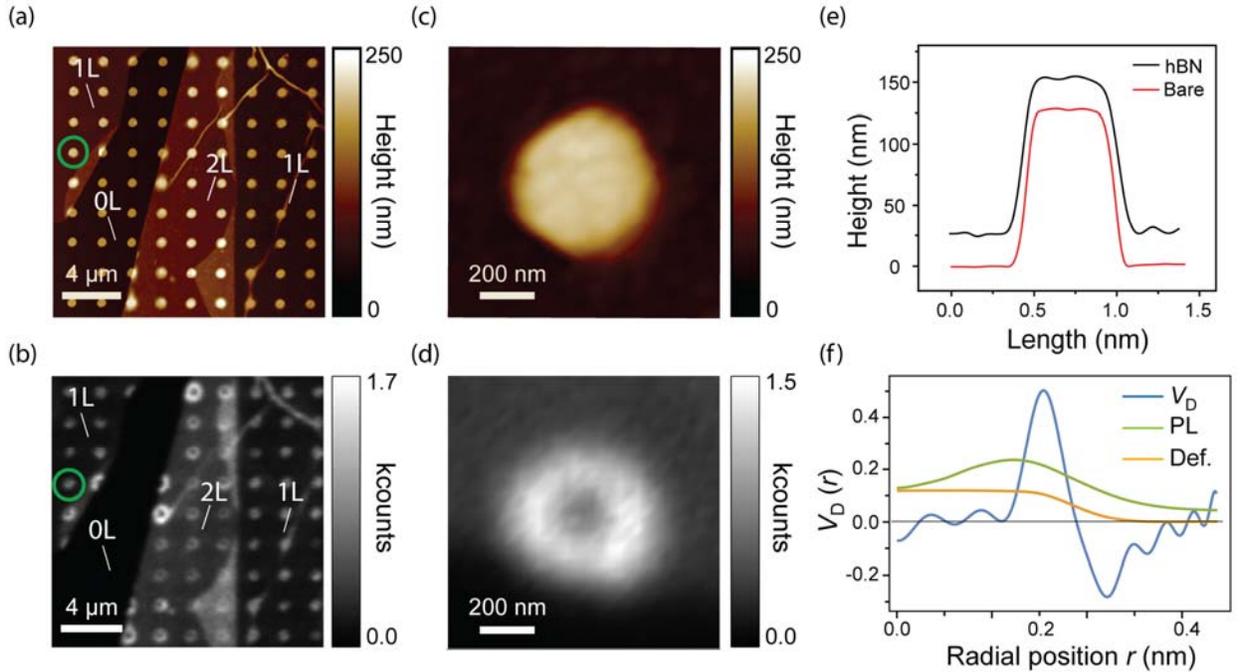

**Figure 4 | Deformation potential in strained hexagonal boron nitride.** (a,b) AFM (a) and confocal (b) microscopy images of a folded hBN flake on a patterned silica substrate. The pillar array is that of Figure 1; 1L and 2L denote one and two hBN layers, respectively; 0L denotes the section of the substrate not covered by hBN. (c,d) Expanded AFM (c) and confocal (d) images of the circled pillar site in (a) and (b), respectively. (e) Cross-sectional view of the AFM image in (c) (solid black trace). Also included for reference is the cross-sectional AFM image of a bare pillar (solid red trace). (f) Calculated deformation potential $V_D(r)$ for holes as a function of the radial distance $r$ to the center of the pillar in (e) (solid blue line). For comparison, the solid green trace is the cross section of the fluorescence pattern in (d) and the solid orange trace is a polynomial fit to the hBN deformation in (e). Both are scaled to arbitrary units. PL: Photo-luminescence; Def: Deformation.

Comparison with the fluorescence pattern from the same pillar site (Figure 4d) shows a reasonable correspondence: Consistent with the KL model, we find a structured, doughnut-shaped emission centered around the pillar. Given the pillar dimensions which is smaller than in Figure 3, the fluorescence dip at the doughnut center is not as pronounced. Interestingly, we find clearer doughnut patterns in the section of the sample where the flake folds on itself to form a second hBN layer. This observation is in qualitative agreement with the wider contours inherent to a multi-layer structure but further work will be needed to gain a fuller understanding on how strain propagates from one layer to the next.

Our findings open interesting opportunities to control defect emission in hBN via the local control of charge. For example, the use of external gate nano-electrodes to create local electrostatic potentials[34] should make it possible to controllably activate emitters on demand. Also intriguing is the use of surface chemistry to control defect emission through selective surface functionalization; this approach should prove useful for sensing applications, e.g., to optically herald molecular binding events on the hBN surface.

**METHODS**

*Substrate and sample preparation.* The silica nano-pillars were made by masked etching of a 300 nm thick thermal oxide on a Si wafer via electron beam lithography. The desired geometry of the $SiO_2$ pillars was first written into a 300 nm layer of negative resist (Ma-N 2403) on top of the thermal oxide wafer via electron beam lithography (Eliox ELS-G100). Then, after development in MIF 726, deep reactive ion etching (DRIE) was utilized to anisotropically etch the patterned wafer. The polymerized resist masks the $SiO_2$ layer from the etchant gas ($CHF3$) and enables the $SiO_2$ pillars to form from the protected silica. Finally, the excess resist is removed via a two-step process; first the majority of the resist is dissolved by the solvent Remover PG and then the pillar substrate is subjected to an $O_2$ plasma for 10 minutes to fully remove any leftover resist. It was found that the O$_2$ plasma etches the silica further ~15 nm, which is attributed to contaminates present on the DRIE chamber walls.

The hBN sample studied herein was purchased from Graphene Supermarket as a 20-nm-thick flake, grown by CVD on a 25-μm-thick Cu substrate. The flake was transferred to the patterned silica wafer by a Poly(methl methacrylate) (PMMA) transfer method[35]. First, a 200 nm layer of PMMA was spin coated onto the h-BN/Cu substrate. After a 90 s prebake at 180 $^0$C, the Cu substrate was removed in a bath of ferric chloride at 60 $^0$C. The hBN/PMMA film was then placed in a Radio Corporation of America (RCA) 2 bath to remove any excess Cu and subsequently in a RCA 1 bath to remove any organic impurities. After rinsing with DI water, the film was lifted from the water bath with a nanopillar sample and allowed to dry. The sample was then heated to 180 $^0$C for 20 minutes to remove any trapped gas and subsequently placed in an acetone bath for 90 minutes at 52 $^0$C to remove the majority of the PMMA film.

The pillar substrates were patterned in arrays with various pillar diameters, which ranged from 200 nm to 2 μm, and varying pitches, from 2 μm to 6 μm. We found that the h-BN was supported by the pillars for pillar heights below ~155 nm while pillars with heights above 155 nm showed evidence the h-BN was pierced by the pillars. We also find that a 2 μm pitch is a sufficient distance to allow the h-BN to drape over the pillar and contact the substrate between pillars. Pillars of other shapes such as triangles and squares were also fabricated.

*Optical measurements*. All photo-luminescent measurements reported herein were collected via a custom-built confocal microscope with an infinity-corrected 50x (.83 numerical objective) Olympus objective. The excitation source had a spot size of 1 μm and varied between two different lasers: a continuous wave (cw) laser operating at 460 nm (Thorlabs L462P1400MM), and a 500 fs pulse fiber laser with a repetition rate of 80 Mhz operating at 488 nm (Toptica FemtoFiber pro TVIS). 500 nm and 550 nm long pass filters (Thorlabs FELO500 and FELO550, respectively) were used to cut off the reflected laser light along the collection arm of the microscope. Correlation measurements were conducted via a free-space Hanbury, Brown and Twist interferometer, where a pair of time-synced (Picoquant – Picoharp 300) APDs (MPD PDM) detected the quantum emission. An 80-20 splitter provided real-time spectral analysis; the 20% arm of the emission was steered into an iHR-320 Horiba spectrometer.

**ACKNOWLEDGEMENTS**
N.P. acknowledges support from CREST IDEALS (NSF-1547830). Z.S., M.D., and V.M.M. acknowledge support from the National Science Foundation (USA) through grants EFRI 2DARE (EFMA-1542863). H.J. and C.A.M acknowledge support from the National Science Foundation (USA) through grants NSF-1619896 and NSF-1401632, and from Research Corporation through a FRED award. M.W.D acknowledges support from the National Science Foundation (Australia) through grant DE170100169. A. A. acknowledges support from the Research Council of Lithuania through grant No. M-ERA.NET-1/2015.

**REFERENCES**

[1] J. L. O'Brien, A. Furusawa, J. Vučković, "Photonic quantum technologies", *Nat. Photonics* **3,** 687–695 (2009).

[2] I. Aharonovich, D. Englund, M. Toth, "Solid-state single-photon emitters", *Nat. Photonics* **10,** 631–641 (2016).

[3] P. Michler, A. Kiraz, C. Becher, W.V. Schoenfeld, P.M. Petroff, L. Zhang, E. Hu, A. Imamoglu, "A Quantum Dot Single-Photon Turnstile Device", *Science* **290,** 2282-2285 (2000).

[4] B. Lounis, W.E. Moerner, "Single photons on demand from a single molecule at room temperature", *Nature* **407,** 491–493 (2000).

[5] V. Acosta, P. Hemmer, "Nitrogen-vacancy centers: Physics and applications", *MRS Bull.* **38,** 127–130 (2013).

[6] L.J. Rogers, K.D. Jahnke, M.H. Metsch, A. Sipahigil, J.M. Binder, T. Teraji, H. Sumiya, J. Isoya, M.D. Lukin, P. Hemmer, F. Jelezko "All-optical initialization, readout, and coherent preparation of single silicon-vacancy spins in diamond", *Phys. Rev. Lett.* **113,** 1–5 (2014).

[7] M.W. Doherty, N.B. Manson, P. Delaney, F. Jelezko, J. Wrachtrup, L.C.L. Hollenberg, "The nitrogen-vacancy colour centre in diamond", *Phys. Rep.* **528,** 1–45 (2013).

[8] D. Riedel, F. Fuchs, H. Kraus, S. Väth, A. Sperlich, V. Dyakonov, A. A. Soltamova, P. G. Baranov, V. A. Ilyin, and G. V. Astakhov, "Resonant addressing and manipulation of silicon vacancy qubits in silicon carbide", *Phys. Rev. Lett.* **109,** 1–5 (2012).

[9] W.F. Koehl, B.B. Buckley, F.J. Heremans, G. Calusine, D.D. Awschalom, "Room temperature coherent control of defect spin qubits in silicon carbide", *Nature* **479,** 84–87 (2011).

[10] Y-M. He, G. Clark, J.R. Schaibley, Y. He, M-C. Chen, Y-J. Wei, X. Ding, Q. Zhang, W. Yao, X. Xu, C-Y. Lu, J-W. Pan, "Single Quantum Emitters in Monolayer Semiconductors", *Nat. Nanotech.* **10,** 497–502 (2015).

[11] C. Chakraborty, L. Kinnischtzke, K.M. Goodfellow, R. Beams, A.N. Vamivakas, "Voltage-controlled quantum light from an atomically thin semiconductor", *Nat. Nanotechnol.* **10,** 507–511 (2015).

[12] A. Srivastava, M. Sidler, A.V. Allain, D.S. Lembke, A. Kis, A. Imamoğlu, "Optically active quantum dots in monolayer WSe$_2$", *Nat. Nanotechnol.* **10,** 491–496 (2015).

[13] M. Koperski, K. Nogajewski, A. Arora, V. Cherkez, P. Mallet, J.-Y. Veuillen, J. Marcus, P. Kossacki, M. Potemski, "Single photon emitters in exfoliated WSe$_2$ structures", *Nat Nano* **10,** 503–506 (2015).

[14] P. Tonndorf, R. Schmidt, R. Schneider, J. Kern, M. Buscema, G.A. Steele, A. Castellanos-Gomez, H.S.J. van der Zant, S. Michaelis de Vasconcellos, R. Bratschitsch, "Single-photon emission from localized excitons in an atomically thin semiconductor", *Optica* **2,** 347–352 (2015).

[15] T.T. Tran, K. Bray, M.J. Ford, M. Toth, I. Aharonovich, "Quantum Emission From Hexagonal Boron Nitride Monolayers", *Nat. Nanotechnol.* **11,** 1–12 (2015).

[16] T.T. Tran, C. Zachreson, A.M. Berhane, K. Bray, R.G. Sandstrom, L.H. Li, T. Taniguchi, K. Watanabe, I. Aharonovich, M. Toth, "Quantum Emission from Defects in Single-Crystalline Hexagonal Boron Nitride", *Phys. Rev. Appl.* **5,** 2–6 (2016).

[17] N.R. Jungwirth, B. Calderon, Y. Ji, M.G. Spencer, M.E. Flatté, G.D. Fuchs, "Temperature Dependence of Wavelength Selectable Zero-Phonon Emission from Single Defects in Hexagonal Boron Nitride", *Nano Lett.* **16,** 6052–6057 (2016).

[18] A.L. Exarhos, D.A. Hopper, R.R. Grote, A. Alkauskas, L.C. Bassett, "Optical Signatures of Quantum Emitters in Suspended Hexagonal Boron Nitride", *ACS Nano* **11,** 3328–3336 (2017).

[19] Z. Shotan, H. Jayakumar, C.R. Considine, M. Mackoit, H. Fedder, J. Wrachtrup, A. Alkauskas, M.W. Doherty, V.M. Menon, C.A. Meriles, "Photoinduced Modification of Single-Photon Emitters in Hexagonal Boron Nitride", *ACS Photonics* **3,** 2490-2496 (2016).

[20] G. Grosso, H. Moon, B. Lienhard, S. Ali, D.K. Efetov, M.M. Furchi, P. Jarillo-Herrero, M.J. Ford, I. Aharonovich, D. Englund, "Tunable and high purity room-temperature single photon emission from atomic defects in hexagonal boron nitride", *arXiv:1611.03515* (2016).

[21] T.T. Tran, C. Elbadawi, D. Totonjian, C.J. Lobo, G. Grosso, H. Moon, D.R. Englund, M.J. Ford, I. Aharonovich, M. Toth, "Robust Multicolor Single Photon Emission from Point Defects in Hexagonal Boron Nitride", *ACS Nano* **10,** 7331–7338 (2016).

[22] N. Chejanovsky, M. Rezai, F. Paolucci, Y. Kim, T. Rendler, W. Rouabeh, F. Fávaro de Oliveira, P. Herlinger, A. Denisenko, S. Yang, I. Gerhardt, A. Finkler, J.H. Smet, J. Wrachtrup, "Structural Attributes and Photodynamics of Visible Spectrum Quantum Emitters in Hexagonal Boron Nitride", *Nano Letters* **16***,* 7037–7045 (2016).

[23] M. Kianinia, B. Regan, S. Abdulkader, T.T. Tran, M.J. Ford, I.Aharonovich, M. Toth, "Robust Solid-State Quantum System Operating at 800 K", *ACS Photonics* **4**, 768–773 (2017).

[24] C. Palacios-Berraquero, D.M. Kara, A.R.-P. Montblanch, M. Barbone, P. Latawiec, D. Yoon, A.K. Ott, M. Loncar, A.C. Ferrari, M. Atatüre, "Large-scale quantum-emitter arrays in atomically thin semiconductors", *Nat. Commun.* **8**, 15093 (2017).

[25] A. Branny, S. Kumar, R. Proux, B.D. Gerardot, "Deterministic strain-induced arrays of quantum emitters in a two-dimensional semiconductor", *Nat. Commun.* **8**, 15053 (2017).

[26] K.K. Kim, A. Hsu, X. Jia, S.M. Kim, Y. Shi, M. Dresselhaus, T. Palacios, J. Kong, "Synthesis and Characterization of Hexagonal Boron Nitride Film as a Dielectric Layer for Graphene Devices", *ACS Nano* **6**, 8583 (2012).

[27] J. Sik, J. Hora, J. Humlicek, "Optical functions of silicon at elevated temperatures", *J. Appl. Phys.* **84**, 6291 (1998).

[28] D. Golla, K. Chattrakun, K. Watanabe, T. Taniguchi, B.J. Leroy, A. Sandhu, "Optical thickness determination of hexagonal boron nitride flakes", *Appl. Phys. Lett.* **102**, 161906 (2013).

[29] S.A. Tawfik, S. Ali, M. Fronzi, M. Kianinia, T.T. Tran, C. Stampfl, I. Aharonovich, M. Toth, M.J. Ford, "First principles investigation of defect emission from hBN", arXiv preprint arXiv:1705.05753.

[30] J. N. Reddy, 2006. *Theory and Analysis of Elastic Plates and Shells*. CRC press.

[31] A. Bosak, J. Serrano, M. Krisch, K. Watanabe, T. Taniguchi, H. Kanda, "Elasticity of Hexagonal Boron Nitride: Inelastic X-Ray Scattering Measurements." *Phys. Rev. B* **73**, 041402 (2006).

[32] J. Wiktor, Julia, A. Pasquarello, "Absolute Deformation Potentials of Two-Dimensional Materials", *Phys. Rev. B* **94**, 245411 (2016).

[33] A. Love and E. Hough. 2013. *A Treatise on the Mathematical Theory of Elasticity*. Cambridge university press.

[34] B. Weber, S. Mahapatra, H. Ryu, S. Lee, A. Fuhrer, T.C.G. Reusch, D.L. Thompson, W.C.T. Lee, G. Klimeck, L.C.L. Hollenberg, M.Y. Simmons, "Ohm's Law Survives to the Atomic Scale", *Science* **335**, 64 (2012).

[35] K.K. Kim, A. Hsu, X. Jia, S.M. Kim, Y. Shi, M. Dresselhaus, T. Palacios, J. Kong, "Synthesis and Characterization of Hexagonal Boron Nitride Film as a Dielectric Layer for Graphene Devices", *ACS Nano* **6**, 8583 (2012).